# CROSS SECTIONS FOR THE ASTROPHYSICAL NEUTRON RADIATIVE CAPTURE ON $^{12}$C AND $^{13}$C NUCLEI

S. B. Dubovichenko[1], N. A. Burkova[2], A. V. Dzhazairov-Kakhramanov[3]

[1] *V.G. Fessenkov Astrophysical Institute "NCSRT" NSA RK, 050020, Almaty, Kazakhstan*
[2] *Al-Farabi Kazakh National University, 050040, av. Al-Farabi 71, Almaty, Kazakhstan*
[3] *Institute of Nuclear physics NNC RK, 050032, str. Ibragimova 1, Almaty, Kazakhstan*

**Abstract.** Within the potential cluster model with Pauli forbidden states the possibility of description of the experimental data on the total radiative n$^{12}$C- and n$^{13}$C-capture cross sections in the astrophysical energy range from 25 meV to 1.0 MeV is presented. The $E$1-transition covers the capture data from the scattering states to the ground one of $^{13}$C and $^{14}$C nuclei. Capture on the three low laying excited states 1/2$^+$, 3/2$^-$ and 5/2$^+$ of $^{13}$C was calculated.

**PACS.** Number(s): 26., 25.20.-x, 24.10.-i, 21.60.-n, 21.60.Gx, 02.

## 1. Introduction

Our last works [1-3] reported the data on the astrophysical S-factors [4, 5] for 10 radiative capture reactions on light nuclei treated on the basis of potential cluster model (PCM) with Pauli forbidden states (FS). Classification of the cluster orbital states in PCM make it possible to define the existence and number of the allowed states (AS) in two-body potentials [6]. The results of the phase shift analysis of the experimental data on the differential cross sections of elastic scattering of corresponding free nuclei are used for the construction of phenomenological two-body interaction potentials for the scattering states in the PCM [7-9].

Potentials for the bound states (BS) of light nuclei in cluster channels are constructed not only on the basis of phase shifts, but under some additional demands. So, they should reproduce the binding energy in corresponding nuclear system as well as some other characteristics of nuclei [1-3]. The choice of PCM for treating of cluster systems and thermonuclear processes at astrophysical energies [4, 10] is caused by the fact, that in lots of atomic nuclei the probability of forming of the nucleon associations (clusters), and the degree of their isolation from each other are comparatively high. This is confirmed by the numerous experimental data and different theoretical calculations implemented over the past fifty years [1, 6, 7, 9, 11, 12].

We continue to study the radiative capture processes on light nuclei, and are treating here n$^{12}$C- and n$^{13}$C-capture reactions at astrophysical energies, that included in the main chain of thermonuclear reactions of primary nucleosynthesis - see, for example, [13, 14].

---

[1] Electronic address: dubovichenko@gmail.com
[2] Electronic address: natali.burkova@gmail.com
[3] Electronic address: albert-j@yandex.ru

$^1$H($n,\gamma$)$^2$H($n,\gamma$)$^3$H($^2$H,$n$)$^4$He($^3$H,$\gamma$)$^7$Li($n,\gamma$)$^8$Li($^4$He,$n$)$^{11}$B($n,\gamma$)$^{12}$B($\beta^-$)$^{12}$C($n,\gamma$)$^{13}$C($n,\gamma$)$^{14}$C($p,\gamma$)$^{15}$N($n,\gamma$)$^{16}$N… .

Apparently these very reactions led finally to the formation of the Sun, stars and our Universe [1, 4]. Available experimental data on the total cross sections, for example, for the n$^{12}$C-reaction are given in [15-20] and may be found also in data bases [21, 22]. Since they do not cover the total energy scale for the processes of the Universe formation, but give the common representation on the behavior of radiative capture cross sections in wide energy range. That is why, it is actual to describe these cross sections within the PCM with FS as it was done for the radiative p$^{12}$C- and p$^{13}$C-capture [23, 24].

Note, recently done phase shifts analysis of new experimental data on p$^{12}$C- and p$^{13}$C-scattering at astrophysical energies [25, 26] made it possible to construct quite unambiguous p$^{13}$C- and p$^{12}$C-potentials. They should not differ essentially from the analogue potentials for n$^{12}$C- and n$^{13}$C-scattering and binding states of $^{13}$C in n$^{12}$C-channel and $^{14}$C in n$^{13}$C-channel.

Further we are following the standpoint that p$^{12}$C and n$^{12}$C channels (as well as p$^{12}$C and n$^{12}$C systems) are the isobar analogues. Actually it is reasonable to examine them in comparative way.

## 2. Model and computer methods

Classification of orbital states for n$^{12}$C- and p$^{12}$C-systems by Young schemes was treated in [23]. It was shown that complete system of 13 nucleons may have the set of Young schemes {1} × {444} = {544} + {4441} [27]. The first of the obtained scheme is compatible with the orbital momentum $L = 0$ only and is forbidden, so far as it could not be five nucleons in the $s$-shell.

The second scheme is allowed and compatible with the angular moments $L = 1$ and 3 defined according the Elliot' rules [27]. State with $L=1$ corresponds to the ground bound allowed state of $^{13}$C nuclei in n$^{12}$C-channel with quantum numbers $J^\pi$, $T= 1/2^-$, 1/2. So, there might be one forbidden bound state in $^2S$- wave potential, and $^2P$- wave should have the allowed state only in n$^{12}$C-channel at the energy -4.94635 MeV [28].

Classification of orbital states for p$^{13}$C-system, and hence n$^{13}$C one by the Young' schemes we did in [24]. So, let us remind shortly that for p$^{13}$C-system within the 1$p$-shell we got {1} × {4441} → {5441} + {4442} [24]. The first of the obtained scheme is compatible with the orbital momentum $L = 1$ only. It is forbidden as five nucleons can not occupy the $s$-shell. The second scheme is allowed and is compatible with the angular moments $L = 0$ and 2 [27].

Thus, restricting by the lowest partial waves we conclude that there is no forbidden state in $^3S_1$-potential, but $^3P$-wave has both one forbidden and one allowed state. The last one appeared at the binding energy -8.1765 MeV of n$^{13}$C-system and corresponds the ground state of $^{14}$C nuclei in this channel with $J^\pi =0^+$ [28].

However, we regard the results on the classification of $^{13}$C and $^{14}$C nuclei by orbital symmetry in n$^{12}$C- and n$^{13}$C-channels as the qualitative ones as there are no complete tables of Young schemes productions for the systems with a number of nucleons more than eight [29], which have been used in earlier similar calculations [2, 7, 30, 31]. At the same time, just on the basis of such classification we succeeded with description of available experimental data on radiative p$^{12}$C- and p$^{13}$C-capture [23, 24].

That is why the given above classification procedure by orbital symmetry was used for the determination of a number of forbidden and allowed states in two-body potentials. Note, as the isospin projection in n$^{13}$C-system $T_z = -1$, then the total isospin $T = 1$, and this is the first cluster system among all treated earlier ones pure by isospin with its maximum value [1-3].

The potential in the Gaussian form as usual [1-3] with a point-like Coulomb term is using for the two-body interaction

$$V(r) = -V_0 \exp(-\alpha r^2). \tag{1}$$

For the calculation of nuclear characteristics additional control on the computing of $^{13}$C and $^{14}$C binding energy in ground state along the finite-difference method (FDM) the variational method (VM) was used. In variational method, the expansion of cluster relative motion wave function (WF) by non-orthogonal Gaussian basis was used, and independent parameter variation was performed [32, 33]. The wave function itself has the form [19]

$$\Phi_L(R) = \frac{\chi_L(R)}{R} = R^L \sum_i C_i \exp(-\beta_i R^2),$$

where $\beta_i$ are the variational parameters and $C_i$ are the expansion coefficients.

The behavior of wave functions for bound states (BS) at large distances was controlled by the asymptotic constant $C_W$ (AC) determining by the Whittaker function of the form [34]

$$\chi_L(R) = \sqrt{2k} C_W W_{-\eta L+1/2}(2kR).$$

Here $\chi_L(R)$ is the numerical wave function of the bound state obtained from the solution of the radial Schrödinger equation and normalized to unity; $W_{-\eta L+1/2}$ is the Whittaker function of the bound state determining the asymptotic behavior of the wave function which is the solution to the same equation without the nuclear potential; i.e., at large distances; $k$ is the wave number determined by the channel binding energy; $\eta$ is the Coulomb parameter; and $L$ is the angular momentum of the bound state.

Total radiative cross sections $\sigma(EJ)$ in case of potential cluster model have the following form -see, for example, [35] or [36]:

$$\sigma_c(EJ, J_f) = \frac{8\pi K e^2}{\hbar^2 q^3} \frac{\mu}{(2S_1+1)(2S_2+1)} \frac{J+1}{J[(2J+1)!!]^2} A_J^2(K) \sum_{L_i, J_i} |P_J(EJ, J_f, J_i) I_J(J_f, J_i)|^2,$$

where for convective electric $EJ(L)$-transitions well known expressions are used [35-37]

$$P_J^2(EJ, J_f, J_i) = \delta_{S_i S_f} \left[(2J+1)(2L_i+1)(2J_i+1)(2J_f+1)\right] (L_i 0 J 0 | L_f 0)^2 \begin{Bmatrix} L_i & S & J_i \\ J_f & J & L_f \end{Bmatrix}^2,$$

$$A_J(K) = K^J \mu^J \left[\frac{Z_1}{m_1^J} + (-1)^J \frac{Z_2}{m_2^J}\right], \qquad I_J(J_f, J_i) = \langle L_f J_f | R^J | L_i J_i \rangle.$$

Here μ – reduced mass of colliding particles; $q$ – wave number in initial channel; $L_f$, $L_i$, $J_f$, $J_i$ –angular ant total momentums in initial ($i$) and final ($f$) channels; $S_1$, $S_2$ – particles spins in initial channel; $m_1$, $m_2$, $Z_1$, $Z_2$ – masses and charges of the particles in initial channel; $S_i$, $S_f$ – total spins in initial and final channels (here $S_i = S_f = S$); $K$, $J$ – wave number and momentum of γ-quantum; $I_J$ – overlapping integral of cluster WF and corresponding radial part of $EJ(L)$-transition operator. Let us note, that in our present calculations as well as in previous we nether used such a characteristic as spectroscopic factor - see, for example, [35], i. e. its value is assumed unit.

We used in our calculations the exact values for particle masses $m_n$=1.00866491597 [38], $m_{13C}$=13.00335502 amu [39], constant $\hbar^2/m_0$ was assumed 41.4686 MeV fm$^2$. For proton capture point like Coulomb potential was used in the form $V_{Coul}$(MeV)=1.439975·$Z_1$·$Z_2$/$r$, where $r$ – relative distance between the particles in initial channel in fermi (fm), $Z$ – charges of the particles in elementary charge units. Coulomb parameter $\eta = \mu·Z_1·Z_2·e^2/(k·\hbar^2)$ was used in the form $\eta = 3.44476·10^{-2}·Z_1·Z_2·\mu/k$, here wave number $k$ in fm$^{-1}$ is defined by the energy $E$ of interacting particles $k^2 = 2\mu E/\hbar^2$.

## 3. Total cross sections for neutron capture on $^{12}$C

Here we are following the declared concept that p$^{12}$C and n$^{12}$C channels are the isobar analogues. So, we are doing their analysis in comparison. Early in [23, 25] the interaction potential for $^2S_{1/2}$-wave in p$^{12}$C-scattering channel was constructed in a way to describe correctly the corresponding partial elastic scattering phase shift which has the pronounced resonance at 0.42 MeV.

On the contrary, in n$^{12}$C-system there are no resonances according [28] up to 1.9 MeV. So, in this channel $^2S_{1/2}$-phase shift should reveal smooth behavior in this energy region. We were unable to find in the literature the results of phase analysis for elastic n$^{12}$C-scattering at the energies below 1.0-1.5 MeV [21, 22], and we suppose they should differ notably from those in p$^{12}$C-scattering channel [26]. That is why for determine of the proper behavior of $^2S$-phase shift the corresponding phase analysis of elastic n$^{12}$C-scattering was done at astrophysical energies, viz. from 50 keV to 1.0 MeV [40]. Experimental measurements of differential elastic scattering cross sections in the energy range from 0.05 up to 2.3 MeV was done in [41]. Results of our analysis for $^2S$-phase shift are presented in Figure 1a by black dots.

In present calculations of the radiative n$^{12}$C-capture dipole electric $E1(L)$-transition corresponding to the convective part of $Q_{JM}(L)$ operator [7, 23, 25] was taken into account. This transition in n$^{12}$C→$^{13}$Cγ process occurs from the doublet scattering $^2S_{1/2}$-state onto $^2P_{1/2}$ ground state of $^{13}$C bound in n$^{12}$C-channel, i.e. the reaction of $^{12}$C(n,γ$_0$)$^{13}$C type is treated. Evaluation of E1-transion from $^2D$ scattering state to ground state (GS) shows the cross section two-three orders magnitude less.

Firstly, a potential for the ground state of $^{13}$C in n$^{12}$C-channel without FS was constructed following the results obtained earlier for p$^{12}$C-system [23]. It should reproduce the binding energy of the ground $^2P_{1/2}$-state of $^{13}$C nucleus in n$^{12}$C-channel equals -4.94635 MeV, as well as value of mean square radius 2.4628(39) fm [28]. The radius of $^{12}$C was taken 2.4702(22) fm [39] comparing 2.4829(19) fm [42]; neutron charge radius is zero, and its mass radius is taken as proton one 0.8775(51) fm [38]. So, for p$^{12}$C-system and n$^{12}$C-channel the following central potential was obtained

$$V_{GS} = 135.685685 \text{ MeV}, \quad \alpha_{GS} = 0.425 \text{ fm}^{-2}. \tag{2}$$

Potential (2) gives the binding energy -4.946350 MeV with accuracy $10^{-6}$ by FDM, mean square charge radius $R_{ch} = 2.48$ fm, and mass radius $R_m = 2.46$ fm. Asymptotic constant turned equals 0.99(1) on the interval 5-13 fm. Error for the AC is defined by averaging over the pointed interval.

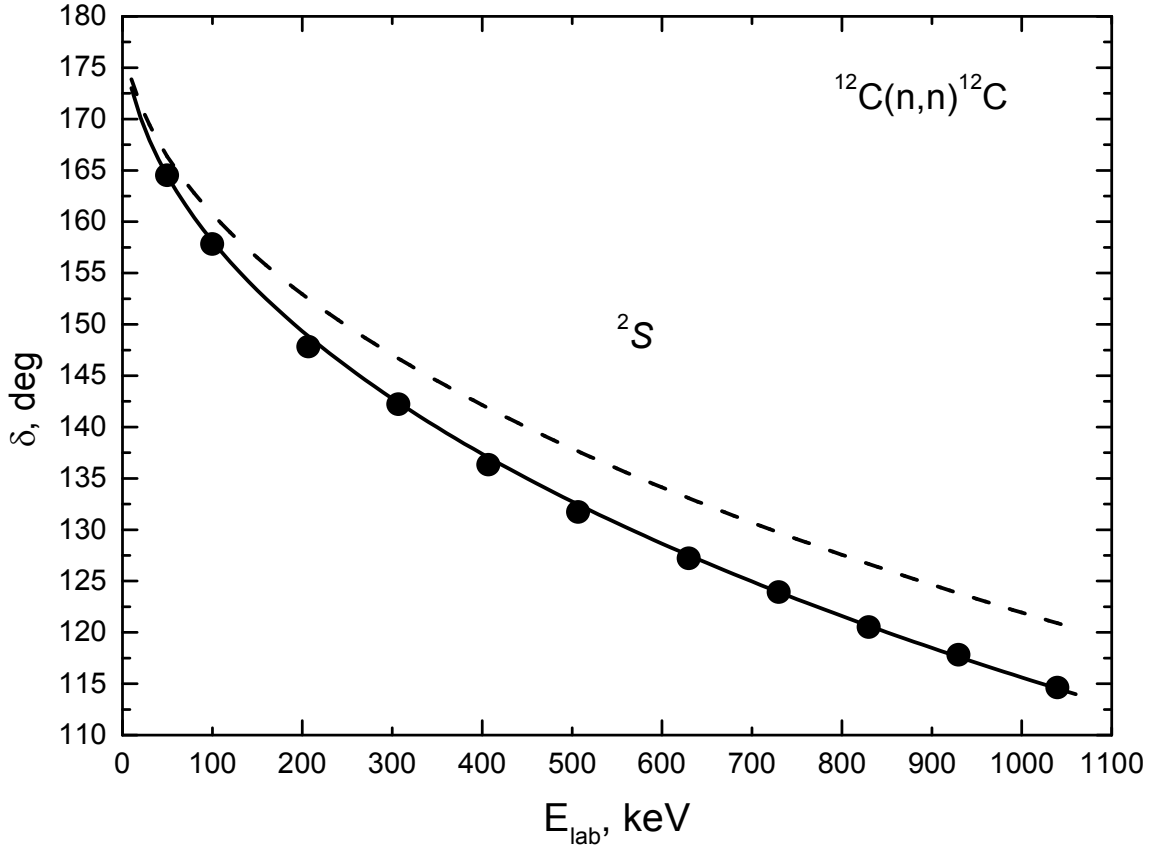

Fig. 1a. Low energy $^2S_{1/2}$-phase shift of elastic n$^{12}$C-scattering. Results of our $^2S$ phase shift analysis – black dots (●) – [40]. Calculations with potential parameters given in text according (4)

Let' note, that according data [43], where the compilation of many results is presented, the obtained value for this constant recalculated with $\sqrt{2k} = 0.942$ to the dimensionless quantity is 1.63(4). According data [14] this value after recalculation is 2.05(18). The redefinition is coming due to another specification for AC differing from our by factor $\sqrt{2k}$

$$\chi_L(R) = C_W W_{-\eta L+1/2}(2kR).$$

There is another set of parameters for n$^{12}$C-potential reproducing the GS of $^{13}$C

$$V_{GS} = 72.173484 \text{ MeV}, \quad \alpha_{GS} = 0.2 \text{ fm}^{-2}. \tag{3}$$

This potential leads to the binding energy -4.94635034 MeV with accuracy $10^{-8}$ by FDM and same charge radius 2.48 fm, but mean square mass radius $R_m = 2.51$ fm is a little bit greater, and AC equals 1.52(1) within the interval 5-18 fm agrees better with data [14, 43]. Solid line in Figure 1b shows WF of such $^2P_{1/2}$ potential.

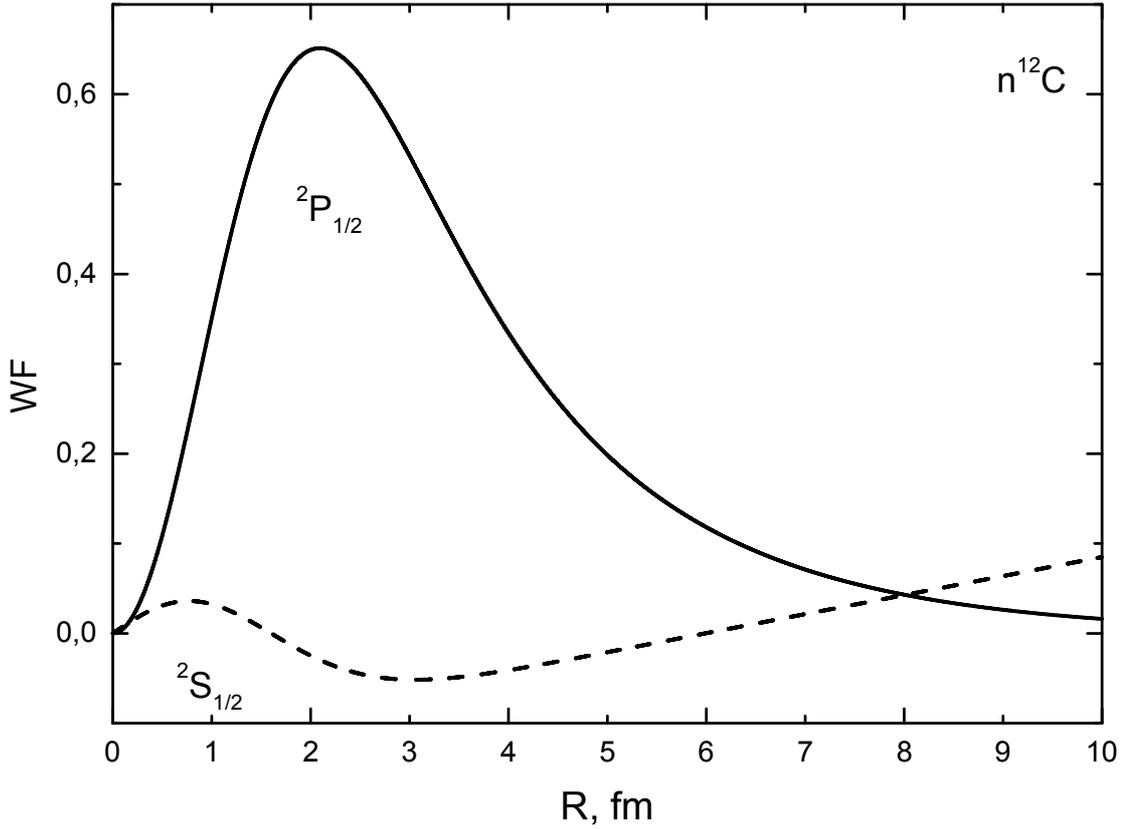

Fig. 1b. Radial wave functions of $^2P_{1/2}$ ground state of $^{13}$C in n$^{12}$C channel and $^2S_{1/2}$ scattering wave at 10 keV. Dots display the integrand of $I_J$ for $E1$ transition.

As the additional computing control for the calculation of binding energy with potential (3) variational method was applied [33]. It gave the energy value - 4.94635032 MeV with dimension $N = 10$ and independent parameter' varying of potential (3). Asymptotic constant $C_W$ of the variational WF is 1.52(2) within the interval 5-15 fm while the residual does not exceed $10^{-12}$ [32]. Charge radius is the same as obtained by FDM.

As the variational energy decreases at the increasing of basis dimension and reaches the upper limit of true binding energy, and the finite-differential energy increases at the reducing of step value and increasing of step number, then it is reasonable to assume the average value for the binding energy - 4.94635033(1) MeV as valid. The accuracy of both methods is $\pm 10 \cdot 10^{-9}$ MeV = $\pm$ 10 meV.

Following the declared isobar-analogue concept, the potential for $^2S_{1/2}$-wave of n$^{12}$C-scattering with parameters obtained for p$^{12}$C-scattering

$$V_S = 102.05 \text{ MeV}, \quad \alpha_S = 0.195 \text{ fm}^{-2},$$

which do not lead to the resonance as it is shown by dash curve in Figure 1a if Coulomb potential is switch off was tried. It gives the total radiative capture cross sections several orders less comparing the experimental data within the treated energy range from 25 meV up to 1 MeV.

Let us turn to the potential describing well $^2S$-phase shift with parameters [40]

$$V_S = 98.57558 \text{ MeV}, \quad \alpha_S = 0.2 \text{ fm}^{-2}. \qquad (4)$$

Phase shift of this potential is given by solid curve in Figure 1a, and dash curve in Figure 1b shows corresponding WF.

The parameters of potential (4) are given with high accuracy for correct description of binding energy in $^2S_{1/2}$ wave laying at -1.856907 MeV towards the threshold of n$^{12}$C channel. Note, if we switch off the Coulomb interaction in initial $^2S_{1/2}$ potential, determined for the p$^{12}$C scattering for correct reproducing of above-threshold resonance at 0.42 MeV, this state becomes bound. So, the potential in n$^{12}$C channel, besides the forbidden, has one allowed bound state corresponding to the first excited state (ES) of $^{13}$C at 3.089 MeV with $J^\pi = 1/2^+$ towards its GS.

Total cross section obtained with (2) for BS and scattering potential (4) is shown by dashed curve in Figure 2a. Calculated cross section is factor two lower than experimental data at 25 meV [15], and it lies a little lower than data [18,19] in the energy range 20-200 keV.

For comparison, let us consider results with the same scattering potential (4), but with GS potential (3), which describes AC correctly. They are shown in Figure 2a by dotted line. It is seen that they lead to correct description of total cross sections obtained in different experimental investigations, beginning from the energy 25 meV to 550 keV. The calculation results for transition from the $^2D_{3/2}$ scattering wave with potential (4) at $L = 2$ and coefficient in cross section for $J_i = 3/2$ for GS of $^{13}$C with potential (3) are shown by dot-dashed line. The scattering phase shift of the $D$ wave is within 1° at the energy less than 1 MeV. The solid line is the sum of dotted and dot-dashed lines, i.e., the sum of transitions $^2S_{1/2} \to {}^2P_{1/2}$ and $^2D_{3/2} \to {}^2P_{1/2}$.

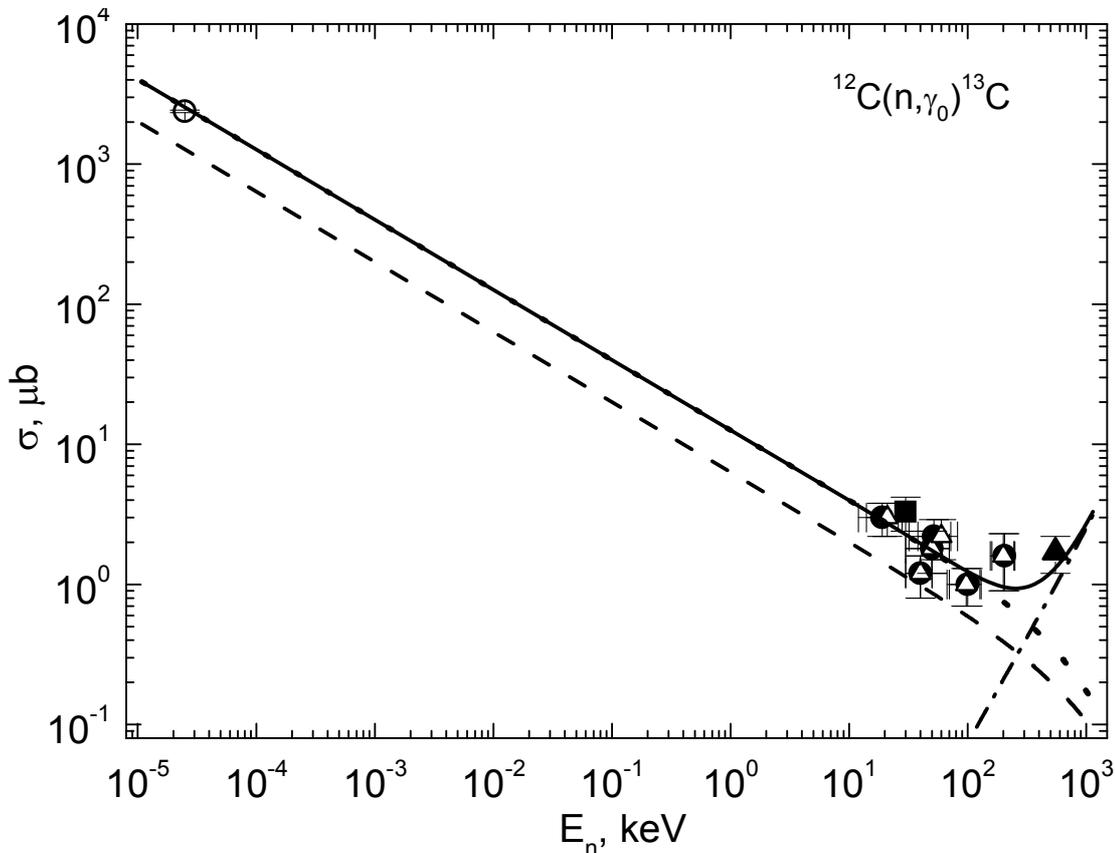

Fig. 2a. The total cross sections of the radiative neutron capture on $^{12}$C at low energies. Experimental data: black squares (■) – [19], black dots (●) – [20], open triangles (Δ) – [18], open circles (○) – [15] and closed triangles (▲) – [16]. Solid line – total cross section calculated with GS potential (2) and scattering potential (4); dash curve – with GS (3) and scattering (4)

We would like to emphasize that these results have been obtained for the potentials (4) and (3) conformed GS characteristics of $^{13}$C, viz. asymptotic constant and low energy n$^{12}$C elastic scattering phase shifts. Thereby, this combination of potentials, describing the characteristics of both discrete and continuous spectra of n$^{12}$C system, allows to reproduce well available experimental data on the radiative neutron capture cross sections for transitions to the GS in the energy range from 25 meV up to 550 keV covering seven orders.

Now treating transitions on to exciting states we want to remark that AC given in [44] for the first ES 1/2$^+$ of $^{13}$C in n$^{12}$C channel is equal to 1.61 fm$^{-1/2}$, and recalculated with $\sqrt{2k} = 0.76$ dimensionless value turned to be 2.12. Besides, in [13] AC equals 1.84(16) fm$^{-1/2}$ was obtained for the first excited state, or recalculated value is 2.42(17).

In the present case $E$1 transition from $P_{1/2}$ and $P_{3/2}$ scattering waves onto $S_{1/2}$ binding excited state in n$^{12}$C channel with (4) is assumed. As $P$ wave has no FS, and there are no negative parity resonances in spectrum of $^{13}$C then corresponding potentials may be regarded zero. While constructing a potential for the binding ES we would orient on the reproducing of the mentioned AC value, as its width affects weakly on the mean square radius.

As a result, the potential (4) with FS was used for excited BS in $S_{1/2}$ wave. It leads to the binding energy of -1.856907 MeV with accuracy 10$^{-6}$ by FDM, charge radius of 2.48 fm, mass radius of 2.67 fm, and AC equals 2.11(1) within the interval 6-24 fm. AC values do not differ too much from results of [13]. Total cross sections given in Figure 2b by solid line display reasonable agreement with experimental data at low energies.

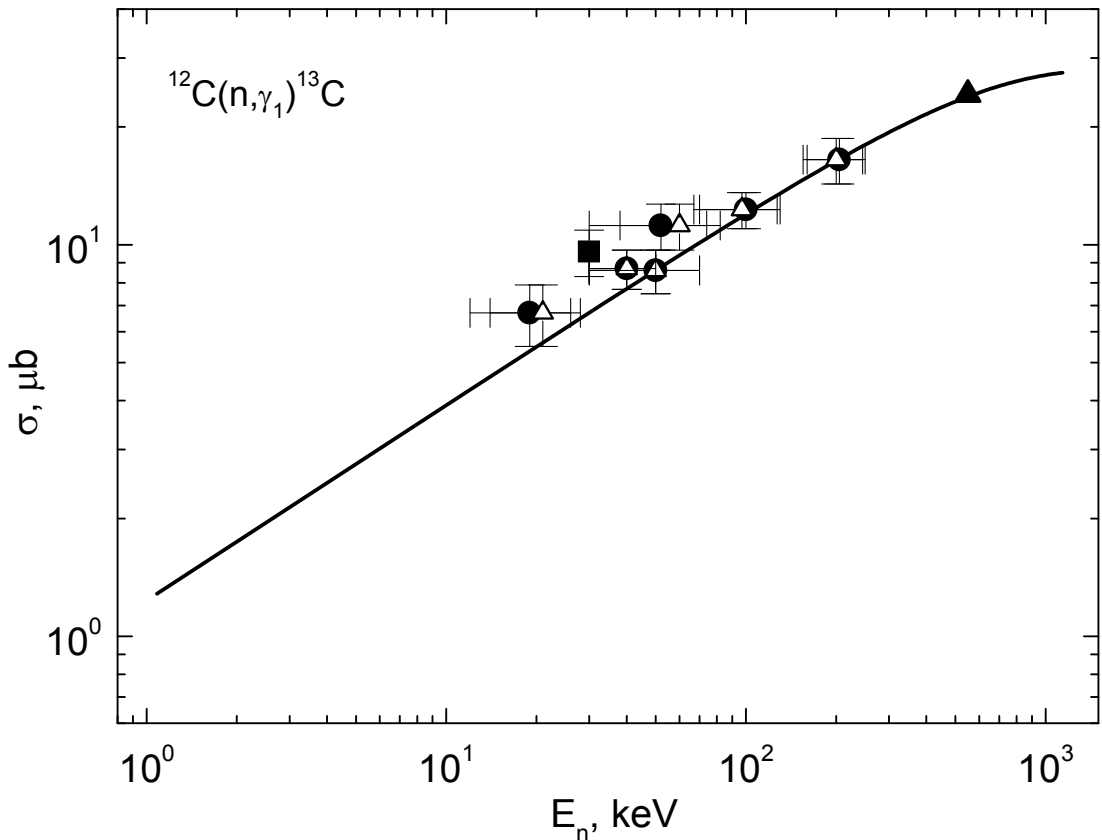

Fig. 2b. The total cross sections of the radiative neutron capture on $^{12}$C on the first excited state 1/2$^+$ of $^{13}$C at low energies. Experimental data: black squares (■) – [19], black dots (●) – [20], open triangles (Δ) – [18], closed triangles (▲) – [16]. Solid line – total cross section calculated with ES potential (4) and scattering $P$ potential with zero depth.

Asymptotic constant for the second excited state 3/2⁻ of $^{13}$C calculated in [44] is 0.23 fm$^{-1/2}$, or 0.33 after recalculation to the dimensionless value with $\sqrt{2k} = 0.69$. For getting the appropriate value of AC, the potential must be very narrow:

$$V_{3/2} = 681.80814 \text{ MeV}, \quad \gamma_{3/2} = 2.5 \text{ fm}^{-2}. \quad (5)$$

This potential gives the binding energy -1.261840 MeV with accuracy $10^{-6}$ by FDM, charge radius 2.47 fm, mass radius 2.44 fm, and AC equals 0.30(1) within the interval 2-24 fm. It does not have FS, and reproduces the AC [44] rather well.

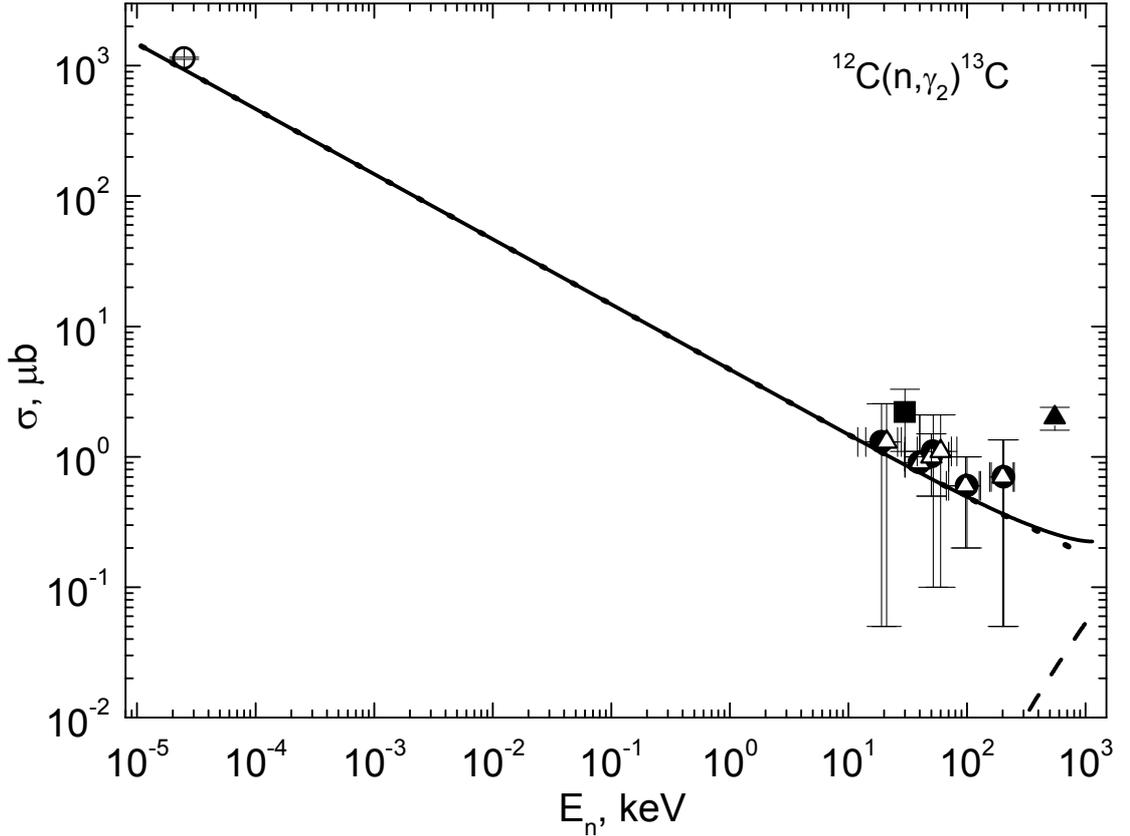

Fig. 2c. The total cross sections of the radiative neutron capture on $^{12}$C on the second excited state 3/2⁻ of $^{13}$C at low energies. Experimental data: black squares (■) – [19], black dots (●) – [20], open triangles (Δ) – [18], open circles (○) – [15] and closed triangles (▲) – [16]. Solid line – total cross section calculated with ES potential (4) and scattering $P_{3/2}$ potential (5).

The calculated cross sections of the neutron capture on $^{12}$C from the $^2S_{1/2}$ scattering state with potential (4) to the $^2P_{3/2}$ level are given in Figure 2c by dotted line together with experimental data [15,18-20]. Dashed line shows the calculation results for the cross section with transition to this ES from the $^2D_{3/2}$ and $^2D_{5/2}$ scattering waves for potential (4) at $L = 2$ and exact coefficients in cross sections for $J_i = 3/2$ and 5/2. Solid line shows the sum of these cross sections. It is well seen that developing approach allows us to obtain acceptable results in description of total cross sections at the transition to the second ES of $^{13}$C. Thereby, the intercluster potentials are conformed to scattering phase shifts as usual and, in the large, correctly describe the main characteristics of the considered BS, which is the second ES of $^{13}$C.

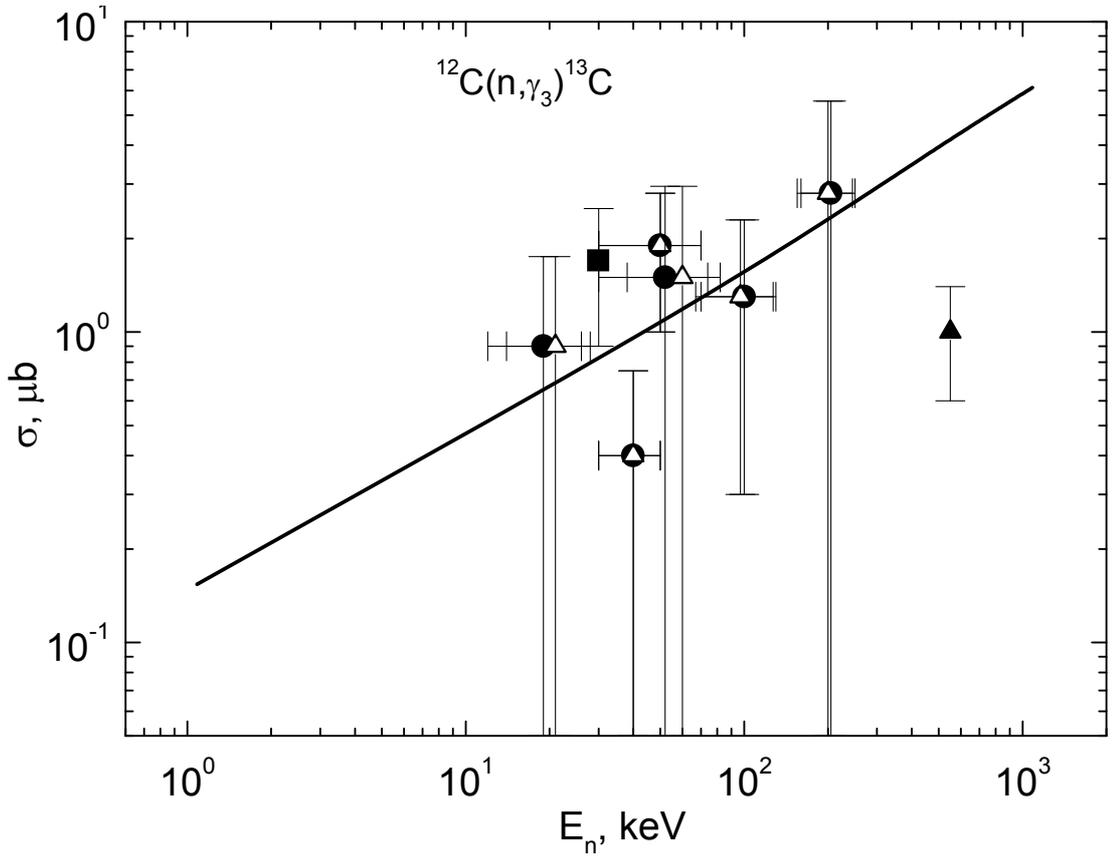

Fig. 2d. The total cross sections of the radiative neutron capture on $^{12}$C on the third excited state $5/2^+$ of $^{13}$C at low energies. Experimental data: black squares (■) – [19], black dots (●) – [20], open triangles (Δ) – [18], closed triangles (▲) – [16]. Solid line – total cross section calculated with ES potential (6) and scattering $P$ potential with zero depth

Let us present the AC for consideration of the transitions from $P_{3/2}$ scattering wave onto binding $D_{5/2}$ state at the energy -1.09254 MeV relatively the threshold of n$^{12}$C channel, which is the third excited state in $^{13}$C. The value 0.11 fm$^{-1/2}$ was obtained in [44], and in [13] it is 0.15(1) fm$^{-1/2}$. Recalculated values at $\sqrt{2k} = 0.665$ turn to be 0.16 and 0.23. For $P_{3/2}$ scattering wave zero potential was used as before. For the binding $D_{5/2}$ state a potential with one FS and same geometry as for the GS of $^{13}$C (3) was used

$$V_D = 263.174386 \text{ MeV}, \quad \gamma_D = 0.2 \text{ fm}^{-2}. \tag{6}$$

It gives the binding energy -1.092540 MeV with accuracy $10^{-6}$ by FDM, charge radius 2.49 fm, mass radius 2.61 fm, and AC equals 0.25(1) within the interval 6-25 fm. It has FS, and reproduces properly the order of magnitude of AC.

The calculated cross sections of the radiative neutron capture on $^{12}$C from the $^2P_{3/2}$ scattering state onto the binding $^2D_{5/2}$ level are shown in Figure 2d by the solid line together with experimental data. Thus, in this case too, the PCM allows us to obtain quite reasonable results in description of total cross sections for the capture to the third ES of $^{13}$C. In addition, the intercluster potentials are agreed with scattering phase shifts as usual and correctly reproduce the main characteristics of each considered BS in n$^{12}$C channel.

As at the energies from $10^{-5}$ to 10 keV the calculated cross section $\sigma_{theor}$ is practically straight line (solid line in Figure 2a) it may be approximated at low energies by simple function

$$\sigma_{ap}(\mu b) = \frac{12.7289}{\sqrt{E_n(keV)}} . \quad (7)$$

Constant value 12.7289 µb keV$^{1/2}$ is defined by one point of cross sections at the minimal energy $10^{-5}$ keV. Modulus of relative deviation between the calculated $\sigma_{theor}$ and approximated $\sigma_{ap}$ cross sections

$$M(E) = |[\sigma_{ap}(E) - \sigma_{theor}(E)]/\sigma_{theor}(E)|$$

in the energy range from $10^{-5}$ up to 10 keV is less than 1.0 %. We would like to assume the same energy dependence shape of the total cross section at lower energies. So, implemented estimation of cross section done at 1 µeV ($10^{-6}$ eV = $10^{-9}$ keV) according (7) resulted 402.5 mb.

## 4. Total cross sections for neutron capture on $^{13}$C

Total cross sections for the radiative capture n$^{13}$C→$^{14}$Cγ process have been calculated with the potential of Gaussian form (1) with zero coulomb term. As the $^3S_1$-wave potential without FS we used firstly the parameters fixed for the p$^{13}$C-scattering channel [24]

$$V_S = 265.4 \text{ MeV}, \quad \alpha_S = 3.0 \text{ fm}^{-2} \quad (8)$$

Figure 3 shows the result of $^3S_1$-phase shift calculation (dashed curve) with p$^{13}$C-potential without Coulomb interaction, but for the scattering n$^{13}$C-channel. It does not reveal now the resonance behavior [25], but depends smoothly from energy. As there is no FS in this system this phase shift starts from zero value.

Potential with one FS of triplet bound $^3P_0$-state should reproduce properly the binding energy of $^{14}$C in $J^\pi = 0^+$ ground state equals in n$^{13}$C-channel -8.1765 MeV [28] as well as describe the mean square radius of $^{14}$C according the experimental value 2.4962(19) fm [28]. The appropriate parameters have been obtained basing on the start set for $^{14}$N in bound p$^{13}$C-state

$$V_{g.s.} = 399.713125 \text{ MeV}, \quad \gamma_{g.s.} = 0.45 \text{ fm}^{-2}. \quad (9)$$

This potential gives the binding energy -8.176500 MeV with FDM accuracy $10^{-6}$ and charge mean square radius $R_{ch} = 2.47$ fm and mass radius 2.47 fm. For the asymptotic constant in dimensionless form [34] value 1.85(1) was obtained in the interval 4-12 fm being averaged by pointed above interval. Note, that the value 1.81(26) fm$^{-1/2}$ was obtained for the AC in this channel in [45], and after the recalculation at $\sqrt{2k} = 1.02$ its dimensionless value 1.77(25) is in agreement with the present calculations.

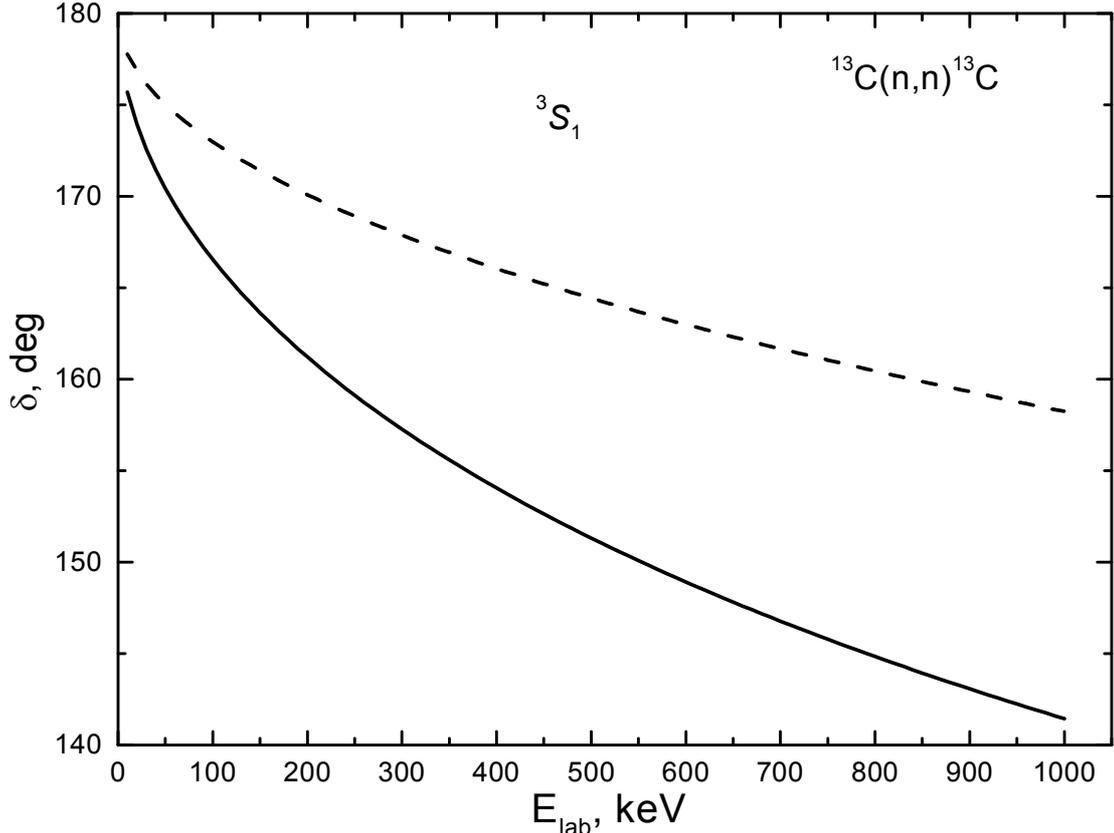

Fig. 3. Low energy $^3S_1$-phase shift of elastic n$^{13}$C-scattering. Dash curve – calculations with potential (8): solid – modified (10)

As the additional computing control for the calculation of binding energy variational method was applied [33]. It gave the energy value -8.176498 MeV with dimension $N = 10$ and independent parameter varying of potential (9). Charge radius and asymptotic constant do not differ from those obtained by FDM. It should be marked that the averaged value for the true binding energy in the discussed potential might be regarded the average value -8.176499(1) MeV obtained by FDM and VM procedures within the computing accuracy ±1.0 eV.

We like to comment that experimental data on the total cross sections on radiative n$^{13}$C-capture given in Figure 4 are taken from [15, 20, 46-48] and were obtained from the Moscow State University data base [21]. Figure 4 shows the pointed experimental data for the energies 25 meV-100 keV.

For the description of total cross sections as in [24] we accounted E1-transition from the non-resonating scattering $^3S_1$- and $^3D_1$-waves obtained with central potential (5) to the bound triplet $^3P_0$-state of $^{14}$C with $J^\pi = 0^+$ in n$^{13}$C-channel generated by potential (9). Calculated total cross sections for $^{13}$C(n, $\gamma_0$)$^{14}$C process at the energies lower 1.0 MeV with defined potential sets overestimate by near two orders the experimental data [20, 46] in the region 10-100 keV.

Experimental data [48] at 25 meV may be reproduced if take the potential depth

$$V_S = 215.77045 \text{ MeV}, \quad \gamma_S = 3.0 \text{ fm}^{-2} \qquad (10)$$

for $^3S_1$-wave at the same geometry. The corresponding phase shift and total cross section are given by solid curves in Figures 3 and 4, respectively. The scattering potential describes properly the bound $^3S_1$ level with $J^\pi = 1^-$ in n$^{13}$C channel but excited at 6.0938 MeV, and leads to the binding energy -2.08270 MeV relatively the threshold, charge and mass radii 2.47 fm, and AC equals 1.13(1) within the interval 2-22 fm. The situation here is similar to those in previous system when the sub-threshold resonance $^3S_1$ state in p$^{13}$C system becomes bound one if the Coulomb interaction is switch off.

Consequently it is seen that slight change of potential depth coordinated with the energy of binding $^3S_1$ level allows to reproduce the experimental data on the total capture cross sections from 25 meV up to 100 keV (see Figure 4). Slowdown of the cross section at 0.5-1.0 MeV is coming due to $E1$ transition from $^3D_1$ scattering wave which input is noticeable in this energy region only. Estimation of $M2$-transition from the resonating $^3P_2$ scattering wave corresponding to $J^\pi = 2^+$ at 141 keV in c.m. to $^3P_0$ ground state shows near 1% input from the $E1$ cross section.

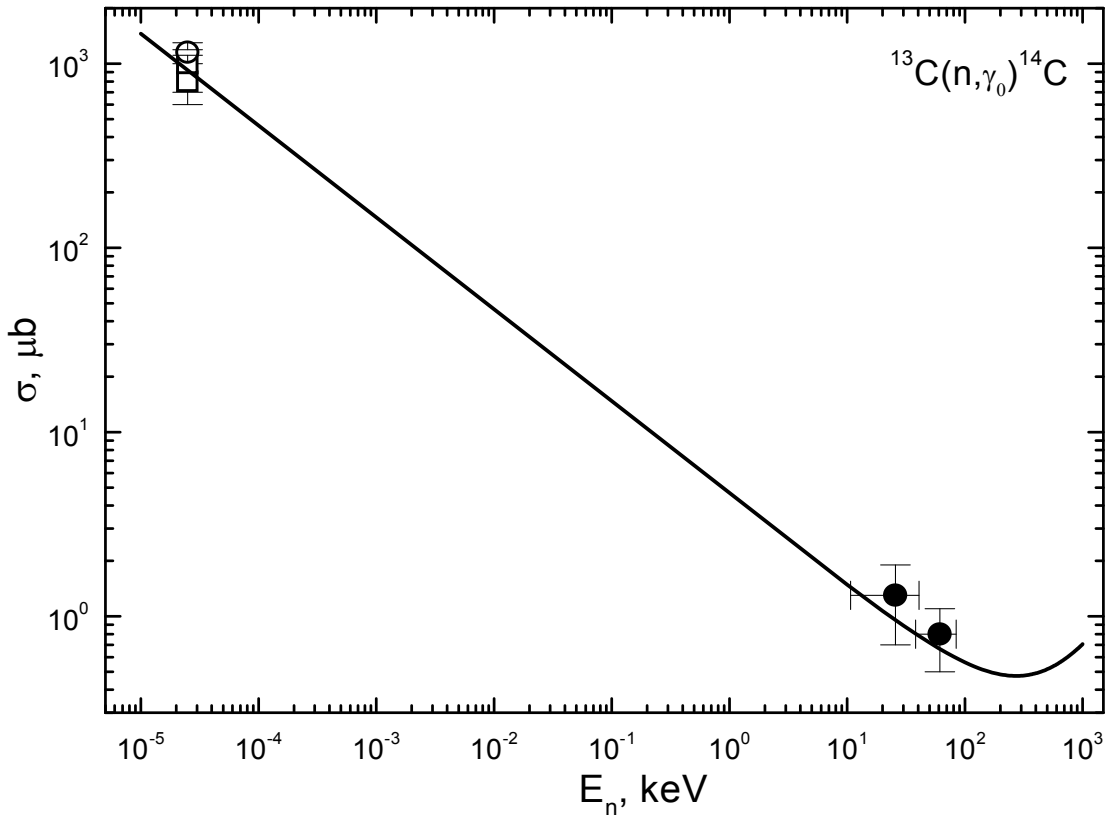

Fig. 4. – Total radiative n$^{13}$C-capture cross sections. Experimental data: black dots (•) – [20], open squares (□) – [47,48], open circles (○) – [15]. Calculations with GS potential (9) and modified scattering potential (10).

Specially should be mentioned that comparing the previous n$^{12}$C system we did not find any independent information on AC for the first ES in binding $^3S_1$ wave. That is why scattering potential (10) may have some ambiguity in parameters. We do not exclude that there might be another set of parameters which may describe correctly the characteristics of bound state, in particular binding energy and total capture cross sections, but leading to somewhat another asymptotic constant.

We did not succeed in search of n$^{13}$C phase shift analysis and experimental data on elastic scattering differential cross sections at the energies below 1.0 MeV. Available data above 1.26 MeV [49] were measured with too large energy step and are not appropriate for the phase shift analysis contrary to the procedure applied successfully for p$^{12}$C- and p$^{13}$C-systems at the energies below 1.0 MeV [25, 40].

It is important to continue our comparative context. For p$^{13}$C-scattering in the resonance region at 0.55 MeV and its width 23(1) keV there have been near 30 measurements of differential cross sections done by several groups at four scattering angles. So detailed data allowed to reproduce within the phase shift analysis [25] the $J^\pi T = 1^-1$ resonance at 8.06 MeV relatively $^{14}$N ground state or 0.551(1) MeV relatively the threshold of p$^{13}$C [28].

In case of n$^{13}$C-scattering data above 1.26 MeV on elastic differential cross sections are given with too large energy step and are not appropriate for the reproducing even the shape of $J^\pi = 1^-$ resonance at 9.8 MeV relatively GS or at 1.75 MeV (l.s.) relatively n$^{13}$C-threshold. Parameters of this resonance are given in Table 14.7 [28]. The situation with information on $^3P_2$-wave of elastic n$^{13}$C-scattering which may be correspond to $J^\pi = 2^+$ resonance at 153 keV (l.s.) with width 3.4 keV (c.m.) is even worse. So, absence of reliable data on scattering differential cross sections at low energies leads to the 10-15% ambiguity for the potential parameters appeared in calculations for $^3S_1$-wave of elastic n$^{13}$C-scattering.

The present situation is analogous to the previous one from the system n$^{12}$C, when changes of the $S$ phase shift for scattering process did not limited by the Coulomb interaction only. The real $S$ phase shift of the n$^{12}$C scattering, obtained in the phase shift analysis and shown in Figure 1a by black dots, has slightly small values than the calculated phase shift for the p$^{12}$C potential with switched off Coulomb interaction, given in Figure 1a by dashed line.

Since in the energy interval from 10 meV up to 10 keV the calculated cross section is practically straight line it may be approximated by simple function like (7) with constant 4.6003 μb keV$^{1/2}$. It was defined by one point of cross section at minimal energy equals 10 meV obtained with parameters set (8) for elastic scattering $^3S_1$-potential and shown in Figure 4. Modulus of relative deviation between the calculated $\sigma_{theor}$ and approximated $\sigma_{ap}$ cross sections defined above as $M(E)$ in the energy range from 25 meV up to 10 keV is less than 0.4%. We would like still to assume the same energy dependence shape of the total cross section at lower energies. So, estimation of cross section done at 1 μeV according (7) gives 145.5 mb.

## 5. Summary and conclusions

Present results show that appropriate n$^{12}$C scattering potential (4) coordinated with corresponding phase shifts [40] together with correct reproducing of $^{13}$C GS enable to describe the available experimental data on the radiative neutron capture cross sections at the energies from 25 meV to 100 keV. All potentials satisfied the classification of FS and AS by orbital Young schemes. Potential constructed for the GS reproduces the basic characteristics of $^{13}$C, i.e. binding energy in n$^{12}$C-channel, mean square route and asymptotic constant.

So, these results may be regarded as one more confirmation of the success of cluster model approach applied early to the radiative neutron processes in other systems [50]. PCM succeeded also in description of radiative capture reactions of protons and other charge clusters on light nuclei [1-3].

Constructed within PCM two-body n$^{13}$C-potentials for $^3S_1$-wave and $^{14}$C GS,

show good results for the total neutron radiative capture on $^{13}$C in the energy range from 25 meV up to 100 keV. Two-body potential used for the bound $n^{13}$C-system reproduces well the basic GS characteristics of $^{14}$C, as well as it was done for $^{14}$N in p$^{13}$C-channel [24].

At a time, it is rather difficult make a certain and final conclusions on potential depth for elastic scattering $^3S_1$-wave as there are essential ambiguity in available experimental radiative capture data. It seems this very value define the calculation results for the radiative neutron cross section capture on $^{13}$C.

New measurements of differential cross sections for elastic $n^{13}$C-scattering in the energy region up to 1.0 MeV with sufficient step might provide the careful phase shift analysis and define the shape of $^3S_1$ elastic phase. This may improve the definition of scattering potential and realize more unambiguous calculations of total radiative n$^{13}$C-capture cross sections.

## Acknowledgments

We would like to express our thanks to Professor R. Yarmukhamedov for the detailed consultations on the asymptotic normalization constants for the treated channels.